\documentclass[12pt,english]{article}
\usepackage[english]{babel}
\usepackage[dvips]{graphicx}
\makeatletter
\usepackage{amssymb}
\usepackage{latexsym}
\usepackage{amsmath}
\usepackage{bbm}
\usepackage{amsthm}

\usepackage{psfrag}

\newtheorem{proposition}{Proposition}[section]

\renewcommand{\Omega}{\mathrm{D}}

\begin{document}
\title{Optimal Base Station Placement: A Stochastic Method Using Interference Gradient In Downlink Case}

\author{
Salman Malik\footnote{INRIA Paris-Rocquencourt, France. Email: \texttt{salman.malik@inria.fr}},
Alonso Silva\footnote{INRIA Paris-Rocquencourt, France. Email: \texttt{alonso.silva@inria.fr}},
Jean-Marc Kelif\footnote{Orange Labs, France. Email: \texttt{jeanmarc.kelif@orange-ftgroup.com}}
}
\date{}

\maketitle

\begin{abstract}
In this paper, we study the optimal placement and optimal number of base stations added to an
existing wireless data network through the interference gradient method.
This proposed method considers a sub-region of the existing wireless data network, hereafter called region of interest.
In this region, the provider wants to increase the network coverage and the users throughput.
In this aim, the provider needs to determine the optimal number of base stations to be added and their optimal placement.

The proposed approach is based on the Delaunay triangulation of the region of interest and
the gradient descent method in each triangle to compute the minimum interference locations.
We quantify the increase of coverage and throughput.
\end{abstract}

\section{Introduction}

Our objective is to increase the coverage and the capacity of a network,
by adding new base stations.
Therefore, we need to find the optimal number and optimal placement
of additional base stations in an already deployed wireless data network.
We consider a sub-region
of the existing wireless data network,
where the service provider
wants to increase the coverage of the network and the throughput of the users.

The problem is to find the
positions of $K$ additional base stations (a variable number to be determined), 
within the region of interest in an existing wireless data network. 
Note that the addition of~$K$ new base stations to the network may impact the 
coverage and the capacity of the existing base stations which makes the problem 
very difficult. 
We show that the problem of finding the optimal location of a set of new base stations when there
is a discrete number of mobile users is an $\mathrm{NP}$-Hard problem.
The problem in its more general form is much harder when
we consider a random distribution of the mobile users,
even for the uniform distribution of the users.
We propose a related problem which is to
find the points of minimum interference in an
existing network.
The optimality is then to
find the minimum interference set of base stations.

We have studied the optimal placement and
optimal number of base stations added to an
existing wireless data network through the
interference gradient method.
Our approach is based on the Delaunay
triangulation of the region of interest (where the triangulation
completely covers the region of interest).
We propose to take as the vertices of this Delaunay triangulation
the positions of the existing base stations.
This is justified by our proof that over each triangle of the considered triangulation there is only one global minimum (over the triangle) of the
interference function. In consequence, there
will be as many candidates of minima over the region
of interest as triangles in the Delaunay triangulation.
The problem is reduced to find over
this set of candidates the minimum interference set of $K$ base stations.
We propose two heuristics to
find this minimum interference points.
Numerical simulations show that the coverage and the
throughput through this method are highly incremented
when our method is considered.

\begin{figure*}
\center
\includegraphics{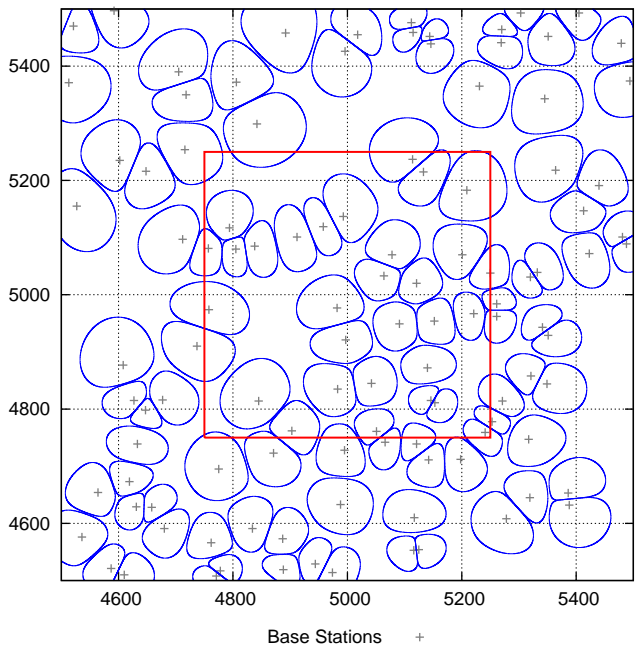}
\caption{$\mathcal{A_R}(\beta)$ of transmitters lying in the central square 
area of $1000\times1000$ square km. Region bounded by a square of side $500$ 
km is the region of interest to place additional base stations. $\beta=1$ and 
$\alpha=4.0$.\label{fig:sigma_default}}
\end{figure*}

\section{Related works}

Plastria~\cite{plastria} presented an overview of the research on locating
one or more new facilities in an environment where
other facilities already exist.
The authors of~\cite{JMK} analyze the coverage and the capacity of a wireless network
regularly distributed.
Gabszewicz and Thisse~\cite{gabszewicz} provided another general survey on the location problem.
Buttazzo and Santambrogio~\cite{buttazzo2005} studied the location problem
but their work does not consider the case when there is an existing wireless data network.
Altman et al.~\cite{AltmanK09} studied the case when there are two providers in the uplink scenario of a cellular network
and the users are placed on a line segment.
Prommak et al.~\cite{Prommak04} design a constraint satisfaction problem
where they consider the transmitted power, frequency bands,
and location constraints.
The authors of~\cite{Keung10} studied a delay-constrained information
coverage which targets an optimal base station placement with the objective
of maximizing information collection within a constraint time,
which is more suitable for wireless sensor networks.

\section{The model}

We consider an existing wireless data network.
In particular, we put ourselves in the context of an LTE (OFDMA-based)
system, where the relevant objective function is to find the optimal
number of base stations and their optimal placement to
maximize the throughput
utility of the data flows.

We consider OFDMA systems supporting fractional frequency reuse (FFR) for interference mitigation
similar to the one used on~\cite{Stolyar2008,Stolyar2009}.
This type of interference mitigation divide frequency and time resources into several resource sets.
Fractional frequency reuse in the context of OFDMA systems has been discussed in cellular network standarization
such as Third Generation Partnership Project (3GPP) and Third Generation Partnership Project 2 (3GPP2)~\cite{3GPP,3GPP2}.

The model is given by the following:
\begin{itemize}
\item Let~${\mathcal S}=\{1,\ldots,S\}$ represent the set of base stations which are randomly distributed 
over the two-dimensional plane according to a Poisson point process (PPP) of intensity~$\lambda$. 
We denote by~$\mathbf{z}_i=(x_i,y_i)$\footnote{We denote the vectors by bold fonts.} the position of base station~$i\in{\cal S}$.
\item and $J$ sub-bands $j\in\mathcal{J}=\{1,\ldots,J\}$ where we denote $W$ the bandwidth of each sub-band.
\item Each sub-band consists of a fixed number~$c$ of sub-carriers.
\end{itemize}
Furthermore, time is divided into slots consisting of a number of OFDMA symbols and transmissions
are scheduled to users by assigning a set of sub-carriers on specific slots.
The time is slotted, so that transmissions within each cell are synchronized,
and do not interfere with each other. To simplify
the exposition we assume that the resource sets
span the entire time period.
Extension to more general resource sets is straightforward~\cite{Stolyar2008,Stolyar2009}.
A transmission in a cell, assigned to a sub-carrier
in a sub-band,
causes interference to only those users in other cells
that are assigned to the same sub-carrier on the corresponding sub-band.
We assume that there is no interference between sub-carriers.

\begin{itemize}
\item We denote by $\tau_{ij}(\mathbf{z})\in [0,1]$ the fraction of time an algorithm
chooses the user located at position~$\mathbf{z}$ in sub-carrier~$i$ in sub-band~$j$.
\item We denote by~$R_{ij}(\mathbf{z})\in~[0,B]$, $B<\infty$ the transmission (nominal) rate
in sub-carrier~$i$ in sub-band~$j$, if the user located at position~$\mathbf{z}$ is chosen.
\end{itemize}
Then the average rate a user located at position~$\mathbf{z}$ actually receives is
\begin{equation}
\mathcal{C}(z)=\sum_{j}\sum_{i}\tau_{i,j}(\mathbf{z}) R_{ij}(\mathbf{z}).
\end{equation}

Each base station may need to allocate a set of sub-carriers and
average power~$P_{ij}^{(k)}(\mathbf{z})$, if the user located at position~$\mathbf{z}$
is to be assigned to sub-carrier~$i$ in sub-band~$j$.

Let us denote by~$G_{ij}^{(k)}(\mathbf{z})$ the propagation gain from cell~$k$
to a user located at position~$\mathbf{z}$ in sub-carrier~$i$ in sub-band~$j$.

Then, we consider the~$\mathrm{SINR}$ (Signal to interference plus noise) function:
\begin{equation}
\mathrm{SINR}_{ij}(\mathbf{z})=\frac{G_{ij}^{(k')}(\mathbf{z})P_{ij}^{(k')}(\mathbf{z})}{N_0+
\sum\limits_{k\in\mathcal{S},k\neq k'}G_{ij}^{(k)}(\mathbf{z})P_{ij}^{(k)}(\mathbf{z})},
\end{equation}
where~$N_0$ is the thermal noise.

We use Shannon formula for the rate
\begin{equation}
\mathcal{C}(\mathbf{z}):=W\log_2(1+\mathrm{SINR}_{ij}(\mathbf{z})).
\end{equation}

Let us consider that the propagation gain from cell~$k$
to a user located at position~$\mathbf{z}$ associated to sub-carrier~$i$ in sub-band~$j$
is given by the channel gain
\begin{equation}\label{gain}
G_{ij}^{(k)}(\mathbf{z}):=\frac{\kappa}{(\sqrt{h^2+d_{ij}^{2}})^\alpha},
\end{equation}
where~$d_{ij}$ is the distance between the base station~$\mathrm{BS}_i$ and the mobile terminal~$\mathrm{MT}_j$,
$h$ is the height of the base station, $\kappa$ is a constant factor,
and~$\alpha>2$ is the path loss exponent.

To simplify the exposition we consider
the distance between base stations
to be much greater that the height of the base station in eq.~\eqref{gain},
and $\kappa$ is constant.
In the following we will drop the sub-indices
and analyze just the interference between sub-carriers.
This scenario is general and it may be applied
to any interference problem.

The objective of this work is to find the optimal positions of $K$ (variable) additional base stations, 
within a bounded region of interest in an existing wireless network, such that
it maximizes the coverage and the capacity of the existing base stations and these $K$ additional base stations.
Note that the addition of~$K$ new base stations to the network may impact the 
coverage and the capacity of the existing base stations which makes the problem 
very difficult. 

The coverage of a base station is given by the $\beta$-$\mathrm{SINR}$ threshold
at which the basic service is provided.
Considering that $P_{ij}^{(k)}$ is constant for all $i,j,k$,
and assuming negligible thermal noise,
we define the $\beta$-reception area of the base station~$i\in{\cal S}$, 
denoted by~$\mathcal{A_R}_i(\beta)$, which represents the area where the transmission from base 
station~$i$ is received with signal to interference plus noise ratio ($\mathrm{SINR}$) greater or equal to~$\beta$,
i.e.,
\begin{equation}
\mathcal{A_R}_i(\beta)=\{\mathbf{z}\,:\, 
\lvert\mathbf{z}-\mathbf{z}_{i}|^{-\alpha}\geq\beta\sum_{j\neq i}|\mathbf{z}-\mathbf{z}_{j}|^{-\alpha}\}
\end{equation}
where $\lvert\mathbf{z}\rvert$ represents the Euclidean norm of the vector \mbox{$\mathbf{z}=(x,y)$},
i.e., $\lvert\mathbf{z}\rvert=\sqrt{x^2+y^2}$.

The coverage of the network is given by the sum over all the base stations of their $\beta$-reception areas
\begin{equation}
\mathrm{Coverage}(\beta)=\sum_{i\in\mathcal{S}}\mathcal{A_R}_i(\beta).
\end{equation}

The~$\beta$-reception area of the base station~$i$, $\mathcal{A_R}_i(\beta)$,
is computed using the gradient $\mathrm{SINR}$ methodology described in
\cite{2010:INRIA-00504081:2}.

\section{Description of our scheme}

%

\subsection{Points of maximum capacity}
Let us assume that the set of~$S$ existing base stations is given by
\begin{equation}
\mathcal{S}=\{1,\ldots,S\}.
\end{equation}
Let us consider the case we want to incorporate a new base station~$i\notin\mathcal{S}$.
Then the set of~$S+1$ base stations will be given by
\begin{equation}
\mathcal{S'}=\mathcal{S}\cup\{i\},
\end{equation}
The $\mathrm{SINR}$ function of a mobile terminal located at position~$\mathbf{z}$
being served by base station~$\mathrm{BS}_k$ with $k\in\mathcal{S'}$ located at position~$\mathbf{z}_k$ is 
\begin{equation}\label{SINRfg}
\mathrm{SINR}_k(z)=\frac{|\mathbf{z}-\mathbf{z}_k|^{-\alpha}}{\sum\limits_{j\in\mathcal{S'},\,j\neq k}|\mathbf{z}-\mathbf{z}_j|^{-\alpha}}=\frac{f}{g},
\end{equation}
where $f=|\mathbf{z}-\mathbf{z}_i|^{-\alpha}$
and $g=\sum\limits_{j\neq i}|\mathbf{z}-\mathbf{z}_j|^{-\alpha}$

We want to compute the maximum of the capacity function obtained by adding a new base station
depending on the
location of the already existing base stations.
The positions of the existing base stations are given,
and we assume that the distribution of the mobile terminals
is uniform,
then the expected capacity function by adding the new base station
is given by
\begin{equation}\label{expectedCapacity}
\mathbf{E}(\mathcal{C}(\mathbf{z}))=\int_{\mathcal{A}}\mathcal{C}(\mathbf{z})\,d\mathbf{z}.
\end{equation}
where
\begin{equation}\label{cap}
\mathcal{C}(\mathbf{z})=\sum_{k\in S'} W\log_2(1+\mathrm{SINR}_k(\mathbf{z})).
\end{equation}

The optima of this maximization problem are found at stationary points, where the first derivative or the gradient of the objective function is zero.
An equation stating that the first derivative equals zero at an interior optimum is sometimes called a ``first-order condition''.

We want to study the behavior of the gradient of eq.~\eqref{expectedCapacity}:
\begin{equation}\label{gradientCapacity}
\nabla\left(\int_{\mathcal{A}}\mathcal{C}(\mathbf{z})\,d\mathbf{z}\right).
\end{equation}
We can rewrite eq.~\eqref{cap} as follows:
\begin{gather}
\sum_{k\in S'} W\log_2(1+\mathrm{SINR}_k(\mathbf{z}))=\\
\sum_{k\in S',k\neq i} W\log_2(1+\mathrm{SINR}_k(\mathbf{z}))+W\log_2(1+\mathrm{SINR_i}(\mathbf{z}))
\end{gather}

Then in the case when there is a large number of base stations the term
\begin{equation}
\nabla\int_{\mathcal{A}}\sum_{k\in S',k\neq i} W\log_2(1+\mathrm{SINR}_k(\mathbf{z}))\,d\mathbf{z}\sim0
\end{equation}

The relevant term in eq.~\eqref{gradientCapacity} is then giving by the following
\begin{equation}
\nabla\int_{\mathcal{A}} W\log_2(1+\mathrm{SINR_i}(\mathbf{z}))\,d\mathbf{z}
\end{equation}

\begin{figure*}
\center
\includegraphics{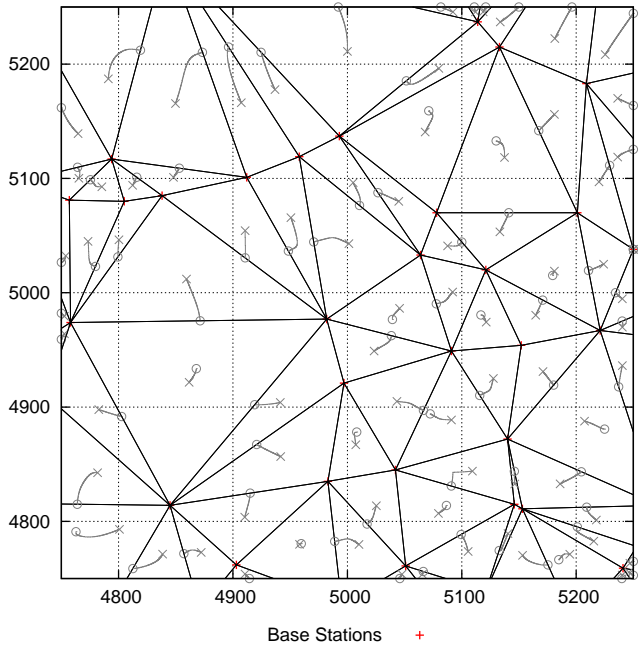}
\caption{Finding points of local minimum interference in the central square area of $500\times500$ 
square km.\label{fig:min_points}}
\end{figure*}

In the low-$\mathrm{SINR}$ regime the problem of maximizing the existing capacity
is equivalent to the problem of maximizing the~$\mathrm{SINR}$.
In the high-$\mathrm{SINR}$ regime both quantities are related
however we do not show a theoretical result.
However, simulation results suggest that even in this
scenario we obtain a good approximation.
In this case, we consider the following function

\begin{equation}
\nabla\int_{\mathcal{A}}\mathrm{SINR}(\mathbf{z})\,d\mathbf{z}
\end{equation}
Notice that we can rewrite eq.~\eqref{SINRfg}
\begin{equation}
\mathrm{SINR}_k(z)=
\frac{|\mathbf{z}-\mathbf{z}_k|^{-\alpha}}{\sum\limits_{j\in\mathcal{S'},\,j\neq k}|\mathbf{z}-\mathbf{z}_j|^{-\alpha}}=\frac{f(\mathbf{z}_k,\mathbf{z})}{g(\mathbf{z})},
\end{equation}
\begin{equation}
\nabla\mathrm{SINR}(\mathbf{z})=\nabla\left(\frac{f}{g}\right)=\frac{g\nabla f-f\nabla g}{g^2}
\end{equation}

In general, this problem is $\mathrm{NP}$-Hard
(consider it as the extension of the coverage problem to the random distribution of users).
Then we propose instead of considering the whole~$\mathrm{SINR}$ function just to consider
$F(\mathbf{z}_k)=\int_{\mathcal{A_R}}f(\mathbf{z}_k,\mathbf{z})\,d\mathbf{z}$ to be a constant function, then in the previous equation we are interested on the behaviour of~$\nabla g$.

The problem of maximizing coverage in wireless data networks with $K$ additional base stations can be 
reduced to the classical maximum coverage problem in computer science~\cite{np-hard}. In maximum 
coverage problem, the inputs are several given sets and a number $K$. These sets may have some 
elements in common. The goal is to select at most $K$ of these sets such that the maximum number of 
elements are covered, {\it i.e.}, the union of the selected sets has maximal size. Formally, the maximum 
coverage problem can be defined as:

{\bf Inputs}: A number $K$ and a collection of sets 
\begin{equation}
S=S_1,S_2,...,S_m
\end{equation}

{\bf Objective}: Find a subset $S'\subseteq S$ of sets, such that $|S'|\leq K$ and the number 
of covered elements $\bigg|\bigcup\limits_{S_{i}\in S}{S_{i}}\bigg|$ is maximized.

The problem of maximum coverage with $N$ given base stations and $K$ additional base stations 
can be reduced to the above described maximum coverage problem. In case of wireless networks, 
the sets $S_i$ represent base stations. The elements of each set may represent the users each base 
station may cover with given SINR threshold $\beta$. Note that these sets may overlap
for $\beta\le1$. For $\beta\ge1$, the zones covered by the base stations do not overlap~\cite{2010:INRIA-00504081:2}.
The location of $K$ additional base stations shall be chosen so that the 
coverage of users, in the network, by $N+K$ base stations is maximum. This is analogous to 
selecting $K$ sets in maximum coverage problem which is an NP-hard problem.

\subsection{Points of minimum interference}

The interference or signal level received at any location $\mathrm{z}$ on the 
two-dimensional plane can be represented by the function 
\begin{equation}
g(\mathbf{z})=\sum_{i\in{\cal S}}\lvert\mathrm{z}-\mathrm{z}_{i}\rvert^{-\alpha}
\end{equation}

There can be many local minima and to identify those points, we have
used the following two-step approach.
\begin{enumerate}
\item We subdivide the region of interest in the network, by using the Delaunay triangulation 
scheme, with the location of the base stations.
\item We use the gradient descent method to locate the point of minimum interference in 
each triangle. Proof of the convexity of the interference function in each triangle is included in 
Proposition~\ref{prop:convexity}. 
\end{enumerate}


\subsection{Delaunay Triangulation}

A Delaunay triangulation for a set of points~$\mathcal S$ in the two-dimensional plane,
denoted by~$\mathrm{DT}(\mathcal S)$ is a triangulation (or subdivision of the two-dimensional plane into triangles)
such that no point in $\mathcal S$ is inside the circumference of any triangle in $\mathrm{DT}(\mathcal S)$.

The Delaunay triangulation of a discrete set of points~$\mathcal S$
corresponds to the dual graph of the Voronoi tessellation for~$\mathcal S$.

We choose the Delaunay triangulation because we have the following property.

\begin{proposition}
Any local minimum interference over the triangle~$\mathcal{T}\setminus\{a_1,a_2,a_3\}$
is also a global minimum interference over the triangle.
\end{proposition}

{\bf Proof.} Consider the function defined in the two-dimensional plane except in~$\mathbf{z}_i$:
\begin{equation}
g(\mathbf{z},\mathbf{z}_i):=\frac{1}{\lvert\mathbf{z}-\mathbf{z}_i\rvert^{\alpha}}
\end{equation}
Then the function of interference is given by
\begin{equation}
g(\mathbf{z})=\sum_{i\in\mathcal{S}}g(\mathbf{z},\mathbf{z}_i)
\end{equation}
Let us consider the triangle of vertices~$a_1, a_2, a_3$ where the three
vertices are the position of three base stations:
\begin{equation}
\mathcal{T}=\{\mathbf{z}\,:\, \mathbf{z}=\sum_{i=1}^3\lambda_i a_i\,;\, 0\le\lambda_i\le1,\,i\in\{1,2,3\},\,\sum_{i=1}^3\lambda_i=1\}
\end{equation}

\begin{proposition}\label{prop:convexity}
The function~$g(\mathbf{z},\mathcal{A_R},\mathbf{z}_i)$ is convex in~$\mathbf{z}$ over \mbox{$\mathcal{T}\setminus\{a_1,a_2,a_3\}$}.
\end{proposition}

{\bf Proof.} We just need to prove that~$\frac{d^2g}{d\mathbf{z}^2}$ is positive.

Let $r_i:=\lvert\mathbf{z}-\mathbf{z}_i\rvert$. The gradient of~$g(\mathbf{z},\mathbf{z}_i)$ in $\mathbf{z}=(x,y)$ is:
\begin{equation}
\frac{dg}{d\mathbf{z}}=-\alpha\lvert\mathbf{z}-\mathbf{z}_i\rvert^{-\alpha-1}\frac{\mathbf{z}-\mathbf{z}_i}{\lvert\mathbf{z}-\mathbf{z}_i\rvert}
=-\alpha\frac{\mathbf{z}-\mathbf{z}_i}{\lvert\mathbf{z}-\mathbf{z}_i\rvert^{\alpha+2}}.
\end{equation}

This means that
\begin{equation}
\frac{dg}{dx}=-\alpha\frac{x-x_i}{r_i^{\alpha+2}}\quad\textrm{and}\quad\frac{dg}{dy}=-\alpha\frac{y-y_i}{r_i^{\alpha+2}}
\end{equation}

\begin{eqnarray}
\frac{d^2g}{d\mathbf{z}^2}&=&
-\alpha\left(\frac{1}{\lvert\mathbf{z}-\mathbf{z}_i\rvert^{\alpha+2}}+
(-\alpha-2)\lvert\mathbf{z}-\mathbf{z}_i\rvert^{-\alpha-3}\frac{(\mathbf{z}-\mathbf{z}_i)^2}{\lvert\mathbf{z}-\mathbf{z}_i\rvert}\right)\nonumber \\
&=&-\alpha\left(\frac{1}{\lvert\mathbf{z}-\mathbf{z}_i\rvert^{\alpha+2}}-(\alpha+2)\frac{(\mathbf{z}-\mathbf{z}_i)^2}{\lvert\mathbf{z}-\mathbf{z}_i\rvert^{\alpha+4}}\right)\nonumber \\
&=&\alpha(\alpha+1)\frac{1}{\lvert\mathbf{z}-\mathbf{z}_i\rvert^{\alpha+2}}
\end{eqnarray}

Since there is no element of~$\mathcal{S}$ included on~$\mathcal{T}\setminus\{a_1,a_2,a_3\}$,
then the function~$g$ is convex on~$\mathcal{T}\setminus\{a_1,a_2,a_3\}$

The sum of convex functions is convex, then over the domain~$\mathcal{T}\setminus\{a_1,a_2,a_3\}$,
the function
\begin{equation}
g(\mathbf{z})=\sum_{i\in\mathcal{S}}g(\mathbf{z},\mathbf{z}_i)
\end{equation}
is convex over the domain~$\mathcal{T}\setminus\{a_1,a_2,a_3\}$.

Any local minimum of a convex function is also a global minimum.

{\bf Important note.} Notice that the fact that we are restricting our domain
to be inside the triangle help us to determine that there is
only one global minimum since the domain is a convex subset.

\subsection{Gradient descent method}

Gradient descent method~\cite{Bertsekas99} is based on the observation that if the real-valued
function is defined and differentiable in a neighborhood of a point
$\mathbf{z}^0$, then the function $g(\mathbf{z})$ decreases fastest if one goes from $\mathbf{z}^0$
in the direction of the negative gradient of $g$ at $\mathbf{z}^0$, $-\nabla g(\mathbf{z}^0)$.
Since our objective is to find the minimum of~$g$, then we propose to use the gradient descent method.
It follows that, if
\begin{equation}
\mathbf{z}^1=\mathbf{z}^0-\delta t\frac{\nabla g(\mathbf{z}^0)}{||\nabla g(\mathbf{z}^0)||}
\end{equation}
and therefore:
\begin{equation}
\mathbf{z}^{n+1}=\mathbf{z}^{n}-\delta t\frac{\nabla g(\mathbf{z}^{n})}{||\nabla g(\mathbf{z}^{n})||}
\end{equation}
where $\delta t>0$ is the step size. Note that $g(\mathbf{z}^{n+1})<g(\mathbf{z}^{n})$. First 
approximate location of minimum interference point, $\mathbf{z}^0$, is the centroid 
of the triangle:
\begin{equation}
\mathbf{z}^0=(x^0,y^0)=\bigg(\frac{x_{p_{1}}+x_{p_{2}}+x_{p_{2}}}{3},\frac{y_{p_{1}}+y_{p_{2}}+y_{p_{3}}}{3}\bigg)
\end{equation}
where $p_{1},p_{2}$ and $p_{3}$ are the vertices of the triangle.
In order to ensure that the points $\mathbf{z}^{n}\,(n\ge0)$ lie inside the triangle,
we have used the method described in \cite{triangle_interior}

\subsection{Heuristics}
Here we propose two heuristics to add $K$ additional base stations at the points of 
minimum interference identified by the gradient descent method~\cite{Bertsekas99}. Let ${\cal P}$ be the 
set of the location of these points of minimum interference.
\begin{enumerate}
\item {\it Heuristic 1:} The optimal position of an additional base station is found as follows.
	\begin{enumerate}
	\item Rank points in the set ${\cal P}$ in the ascending order of interference or signal 
	level received from all base stations in the set ${\cal S}$.
	\item Select the lowest ranked element $p$ from the set ${\cal P}$. The location of 
	$p$ shall become the position of the additional base station.
	\item Remove $p$ from the set ${\cal P}$, {\it i.e.}, ${\cal P}:={\cal P}-p$.
	\item For additional base stations, repeat from step (b). 
	\end{enumerate}
\item {\it Heuristic 2:} The addition of a base station at minimum interference point in set ${\cal P}$ 
may increase the interference at other points of set ${\cal P}$ located in the immediate vicinity. 
Therefore, remaining points in set ${\cal P}$, \{${\cal P}-p$\}, should be re-ranked. 
The Delaunay triangulation in this case will be provided by
the Bowyer-Watson algorithm which gives another approach for incremental construction.
It gives an alternative to edge flipping for computing the Delaunay triangles containing a newly inserted vertex.

Steps (a) and (b) of {\it Heuristic 2} are the same as in {\it Heuristic 1}. After addition of a 
base station, the region of interest is re-triangulated to identify new points of minimum 
interference. In other words, the points of minimum interference in the region of interest shall be 
identified before the addition of all new base stations.

\end{enumerate} 

\begin{figure*}
\center
\includegraphics{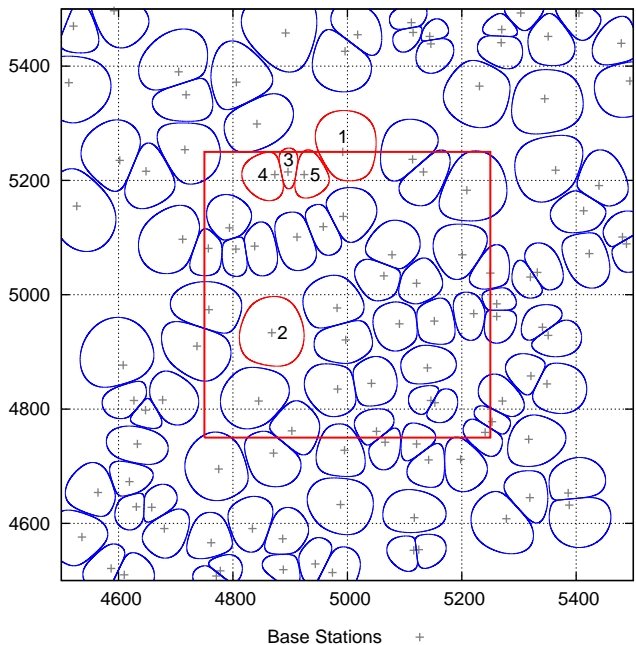}
\caption{$5$ base stations, denoted $1$ to $5$, are added in the region bounded by a square 
of side $500$ km using {\it Heuristic 1}. The number indicates the order of addition to 
the network. $\beta=1$ and $\alpha=4.0$.\label{fig:sigma_1}}
\end{figure*}

\section{Numerical Simulations}

We consider a wireless data network over an area of $10000\times10000$
square km and base stations distributed according to a Poisson point 
process (PPP) of intensity $\lambda=1$ base station per unit square mile. 
We consider the path loss exponent~$\alpha=4$.

In order to ignore edge effects, while computing minimum interference points, we 
assume that our region of interest lies in the center of this network and is a square 
region with each side of length $500$ km. 

Figure~\ref{fig:sigma_default} shows the
$\beta$-reception areas of the base stations distributed according to a Poisson 
point process of intensity $\lambda=1$. It also shows the region of interest marked by a thick lines inside
the network. 

Figure \ref{fig:min_points} shows the process of identification of the points 
of minimum interference in this region of interest. Locations marked by 'x', show 
the first approximation of the points of minimum interference. Gradient descent 
method uses the first approximate position to arrive at the final location, marked 
by 'o', which is the true point of minimum interference in the region formed by 
the Delaunay triangle. Note that, in cases where a triangular region lies partially 
within the region of interest, we are interested in identifying the point of minimum 
interference lying within the region of interest.

\begin{figure*}
\center
\includegraphics{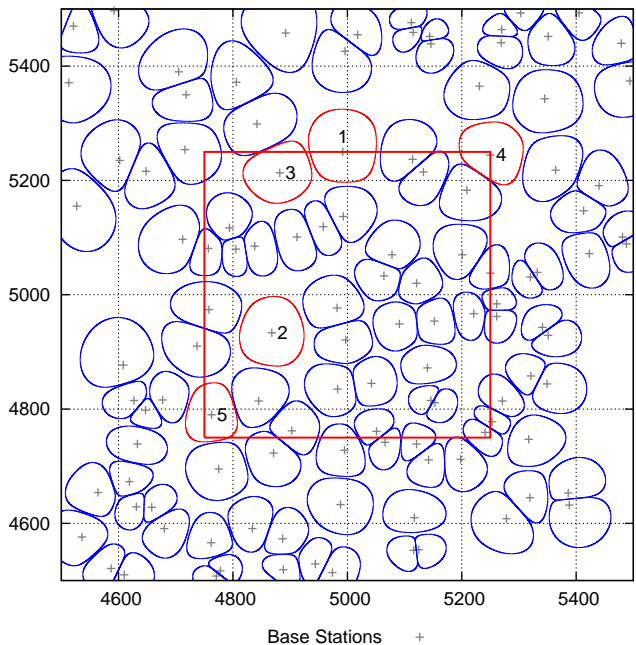}
\caption{$5$ base stations, shown in red, are added in the region bounded by a square 
of side $500$ km using {\it Heuristic~2}. The number indicates the order of addition to 
the network. $\beta=1$ and $\alpha=4.0$.\label{fig:sigma_2}}
\end{figure*}

Figures \ref{fig:sigma_1} and \ref{fig:sigma_2} shows the positioning of $5$ 
additional base stations in the network by using the approaches of {\it Heuristic~1} 
and {\it Heuristic~2} respectively. In order to show the improvement achieved by 
the two heuristics, we have computed the total area covered by the base stations 
with SINR at least equal to $\beta$, {\it i.e.}, 
$\sum_{i\in{\cal S'}}\mathcal{A_R}_i(\beta)$ where ${\cal S'}$ is the set of 
base stations, also including the additional ones in case of {\it Heuristic 1} and 
{\it Heuristic 2}, lying in the region of interest.

Scenario~$0$ indicates the existing wireless data network to which
we want to increase its coverage and its capacity.

Table~1 summarizes the results obtained by both heuristics.
We compute the capacity over the network under three different scenarios:
Scenario~$0$ which indicates the capacity on the existing wireless data network,
and {\it Heuristic 1} and {\it Heuristic 2} explained above.

Table~2 summarizes the results obtained by both heuristics.
We compute the coverage of the network under the three different scenarios described above.



\begin{table}
\begin{center}
\begin{tabular}{|l|c|c|c|}
\hline
 & \textit{Scenario~0} & \textit{Heuristic 1} & \textit{Heuristic 2}\\
\hline
\hline
Capacity & 1.3559 & 1.5614 & 1.7007\\
\hline
Percentage Increase & \- & 15.15\% & 25.42\%\\
\hline
\end{tabular}
\caption{\sl Average capacity over the surface $[\textrm{bits}/s/\textrm{Hz}/\textrm{km}^2]$ over the network and percentage increase of the capacity under different heuristics}
\end{center}\label{table2}
\end{table}

\begin{table}
\begin{center}
\begin{tabular}{|l|c|c|c|}
\hline
 & \textit{Scenario~0} & \textit{Heuristic 1} & \textit{Heuristic 2}\\
\hline
\hline
Total coverage area & 168776 & 190501 & 204628\\
\hline
Coverage percentage & 67,51\% & 76,20\% & 81,85\%\\
\hline
Percentage Increase & \- & 12.87\% & 21.25\%\\
\hline
\end{tabular}
\caption{\sl Coverage of the network (in square km) and percentage increase of the coverage under different heuristics}
\end{center}\label{table1}
\end{table}

\section{Conclusions}
We have studied the optimal placement and
optimal number of base stations added to an
existing wireless data network through the
interference gradient method.
The problem of finding the optimal location of a set of new base stations when there
is a discrete number of mobile users is $\mathrm{NP}$-Hard.
The problem considered is even harder when
we consider a random distribution of the mobile users
even for the uniform distribution.
Our proposed method
considers a sub-region
of the existing wireless data network
(hereafter called region of interest),
where the service provider
wants to increase the coverage and throughput of the users.
Our approach is based on the Delaunay
triangulation of the region of interest (this triangulation
completely covers the region of interest),
and it takes as the vertices of this triangulation
the positions of the existing base stations.
We prove that over each triangle there is only one global minimum of the
interference function. In consequence, there
will be as many candidates of minima over the region
of interest as triangles.
Then the problem is reduced to find over
this set the minimum interference set.
We propose two heuristics to
find the minimum interference points.
Numerical simulations show that the coverage and the
throughput through this method are highly incremented.

\section{Perspectives}
There are many possible extensions to our work,
but we think that the more interesting ones are the following:
\begin{itemize}
\item If we want to take into consideration a more realistic LTE (OFDMA) system, the
mobile user may receive a set of sub-carriers.
\item
We have neglected the fact that the distribution of the users
may be non-homogeneous over the region of interest.
\item
In a future work,
the shadowing fluctuations over the network
will be taken into account.
\end{itemize}


%
%
\bibliographystyle{hieeetr}
\bibliography{sigproc}  
\end{document}